# Implementation Strategies for Multidimensional Spreadsheets

*Paul Mireault*
*Founder, SSMI International*
*Honorary Professor, HEC Montréal*
*Paul.Mireault@SSMI.International*

**ABSTRACT**

Seasoned Excel developers were invited to participate in a challenge to implement a spreadsheet with multi-dimensional variables. We analyzed their spreadsheet to see the different implement strategies employed. We identified two strategies: most participants used a projection of three or four-dimensional variables on the two-dimensional plane used by Excel. A few participants used a database approach where the multi-dimensional variables are presented in the form of a dataset table with the appropriate primary key. This approach leads to simpler formulas.

## 1 Introduction

Expert guidelines for developing spreadsheets, such as (Raffensperger, 2003) and (FAST Standard Organisation, 2016), recommend shorter formulas with fewer references.

Researchers have studied the complexity of spreadsheets because of its impact on the risk of having errors. (Bregar, 2004) developed a complexity metric that takes numerous factors into account, such as the number of operators and operands, the nesting, and the dispersion of references. Combining approaches from Software Engineering and Linguistics, (Reschenhofer, Waltl, & Matthes, 2016) have also developed a metric based the analysis of a spreadsheet's formulas. Using the concept of *smell* in Software Engineering, (Hermans, Pinzger, & van Deursen, 2014)measured complexity with the number of operators and of references, the length of calculation chains and the presence of duplicated formulas.

What all those research projects and expert recommendations fail to do is take into account the complexity of the problem itself, and that cannot be done with only an analysis of the formulas in a spreadsheet file.

Even though a spreadsheet has two dimensions, rows and columns, one of the spreadsheet's dimension is usually used to represent variables. This leaves the other spreadsheet's dimension to represent one of the model's dimension, such as time. But real-world problems can be more complex. For example, we may want to model unit sales by month, by region and by product, thus needing a three-dimensional variable. The challenge's problem had variables in up to four dimensions: month, sector, product and region.

In this paper, we will describe how some Excel users implement a multidimensional structure. The structure is the one described in (Mireault, 2018).

We recruited the participants in the study through an Internet challenge. The challenge kit consisted of the problem description and its solution, presented in the form of a Formula Diagram and a Formula List. The kit also contained an Excel file with the data and the desired Interface sheet and it also contained two screen captures showing the desired results given two different inputs. The participants could then validate themselves to make sure that their calculations were correct.

This research is an explorative study to see what techniques Excel developers use to implement multidimensional variables. Since participation was voluntary, we cannot do any statistical inference with the results.

## 2 Recruiting the participants

An invitation to participate in the challenge was sent to the EuSpRIG mailing list(EuSpRIG, 2019) and on the LinkedIn Excel Developers group(LinkedIn, 2019), with encouragements to redistribute the invitation in other circles. The challenge lasted one month from November 1 to November 31, 2018. The participants had to request the kit by email, which allowed us to measure the interest level. There were 109 kit requests and 17 of them submitted a spreadsheet. We did not investigate why potential participants did not complete the challenge, but some did communicate that they were too busy or that they found the problem too complicated and did not know how to implement it.

The participants were given a problem along with its solution: all the variables and the formulas were presented in the Formula List. Their task was not to solve the problem, but to implement its given solution. The participants were free to use any feature that did not require non-standard features: it had to work in a simple Excel installation.

The Formula List contains 5 variables with more than two dimensions. `MSP Unit Sales` and `MSP Sales Amount` are in the (*Month*, *Sector*, *Product*) dimension set, `MSPR Unit Sales` and `MSPR Variable Cost` are in the (*Month*, *Sector*, *Product*, *Region*) dimension set, and `MPR Unit Sales` is in the (*Month*, *Product*, *Region*) dimension set. Those are the variables that interested us in this analysis.

## 3 Initial analysis

Given the voluntary lack of directives, some participants did not implement all the variables of the Formula List. They avoided implementing some multidimensional variables by building more complex formulas. For example, the two four-dimensional variables, `MSPR Unit Sales` and `MSPR Variable Cost` can be avoided by changing the given formulas for the variables that depend on them: `MPR Unit Sales`, `Monthly Unit Sales` and `Monthly Variable Cost`. So, the formulas

```
MPR Unit Sales = SUM( MSPR Unit Sales )
```

and

```
MSPR Unit Sales = MSP Unit Sales * Region Sales Distribution per Sector
```

were combined by some participants to calculate

```
MPR Unit Sales = SUM( MSP Unit Sales * Region Sales Distribution per Sector ).
```

This was unfortunate for this study, as we were interested in seeing how the variable `MSPR Unit Sales` was implemented in their worksheet. But since the four-dimensional variables were not presented as output variables in the Interface sheet their implementation was mathematically correct.

## 4 Implementation strategies

The implementation strategies used by the participants can be assigned to two broad categories that we'll call the Database approach (DB) and the Variables and Formulas approach (VF). The VF approach is well used in one or two-dimensional spreadsheets in business and finance applications.

In a one-dimensional spreadsheet, the variables, like Unit Sales, Cost and Revenues, are usually represented in rows, with columns on the left-side documenting them with their name and, sometimes, their units. One dimension, like Month or Region, is presented in a row at the top and columns represent the instances of the dimension, such as Jan, Feb, Mar or South, North. The formulas appear in a rectangular shaped area below the dimension row and to the right of the variables. Normally, the formulas linking the variables are written for one instance of the dimension and then copied under the other instances to the right.

Implementing two dimensions starts to be a bit more involved. One method consists of presenting the second dimension along with the first one. All the instances of the first dimension are then repeated for each instance of the second dimension. Thus, if we have twelve months and three regions, the set of twelve months (Jan, Feb… Dec) will be repeated for each of the three regions, giving 36 columns.

Another method consists of subsuming a dimension with variables. Thus, instead of having one variable `Sales` with 36 instances— `Sales(North, Jan)`, to `Sales(East, Dec)`— we create three variables named `Sales North`, `Sales South` and `Sales East`. And there remains only one dimension, Month, with 12 instances. Usually, calculations with variables with the same suffix are grouped together. Thus the calculations of Sales North, Cost North and Profit North would appear in the same area and then would be reproduced and adjusted for Sales East, Cost East and Profit East.

### 4.1 Three-dimensional variables

We had two 3-dimensional variables, `MSP Unit Sales` and `MSP Sales Amount` in the (*Month*, *Sector*, *Product*) dimension set, and `MPR Unit Sales` is in the (*Month*, *Product*, *Region*) dimension set. This variable was also shown as an output variable in the Interface sheet given to the participants, as shown in Figure 1.

|  |  |  | Jan | Feb | Mar | Apr | May | Jun | Jul | Aug | Sep | Oct | Nov | Dec |
|---|---|---|---|---|---|---|---|---|---|---|---|---|---|---|
| Unit Sales by Month, Product and Region | Product S | Regions N | 161 | 182 | 194 | 199 | 211 | 189 | 171 | 148 | 131 | 122 | 128 | 143 |
|  |  | SE | 113 | 125 | 129 | 128 | 133 | 118 | 109 | 97 | 90 | 87 | 93 | 103 |
|  |  | SW | 145 | 169 | 182 | 190 | 208 | 194 | 177 | 151 | 130 | 119 | 118 | 128 |
|  |  | E | 128 | 141 | 145 | 143 | 147 | 130 | 120 | 108 | 101 | 99 | 106 | 118 |
|  |  | W | 93 | 103 | 109 | 109 | 113 | 100 | 91 | 80 | 73 | 70 | 75 | 84 |
|  | Product D | Regions N | 164 | 171 | 173 | 177 | 180 | 149 | 135 | 120 | 118 | 108 | 127 | 145 |
|  |  | SE | 107 | 108 | 103 | 96 | 93 | 75 | 72 | 70 | 75 | 76 | 90 | 101 |
|  |  | SW | 109 | 115 | 114 | 112 | 114 | 97 | 90 | 82 | 82 | 78 | 89 | 100 |
|  |  | E | 111 | 110 | 102 | 89 | 84 | 66 | 67 | 68 | 77 | 82 | 97 | 108 |
|  |  | W | 89 | 91 | 90 | 87 | 86 | 70 | 65 | 61 | 63 | 61 | 72 | 82 |

*Figure 1 MPR Unit Sales in the Interface sheet*

Not surprisingly, all but one participant adopted that representation in their calculation sheet. Since participants had different ways of showing a particular structure (with different formats, spacing, etc.), we will represent this structure with a diagram as shown in Figure 2. The diagram shows that the Month dimension is in columns and the Product and Region dimensions are in row. Furthermore, the Products are in sequence and the Regions are repeated for each instance of the Products. The body of the structure is the calculation of the MPR Unit Sales variable.

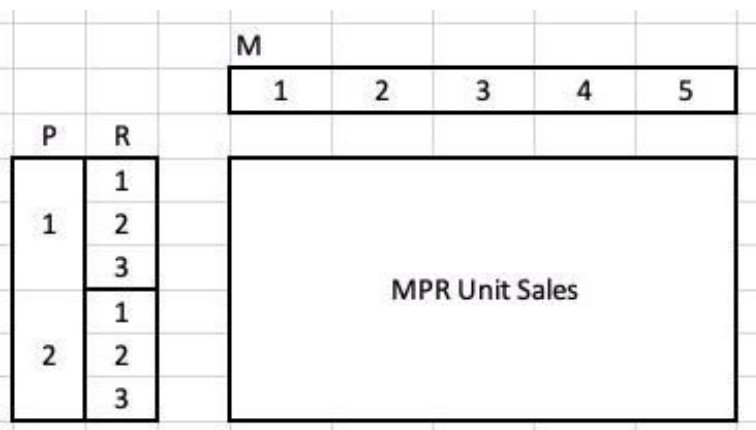

*Figure 2 Diagram of Structure MPR1*

One participant chose to calculate the variable with a different structure. In this case, all the dimensions were represented in rows, as shown in Figure 3. Our representation of that structure is shown in Figure 4.

| Month | Region | Product | MPR Unit Sales |
|---|---|---|---|
| Jan | N | Standard | 160.9580294 |
| Feb | N | Standard | 181.642437 |
| Mar | N | Standard | 193.6926816 |
| Apr | N | Standard | 199.2323309 |
| May | N | Standard | 211.0830231 |
| Jun | N | Standard | 189.4979927 |
| Jul | N | Standard | 171.0720907 |
| Aug | N | Standard | 148.0645629 |
| Sep | N | Standard | 131.4326926 |
| Oct | N | Standard | 121.840506 |
| Nov | N | Standard | 127.6870749 |
| Dec | N | Standard | 142.8964478 |
| Jan | SE | Standard | 113.4317641 |
| Feb | SE | Standard | 124.5724551 |
| Mar | SE | Standard | 129.4009003 |
| Apr | SE | Standard | 128.4498505 |
| May | SE | Standard | 133.2302664 |
| Jun | SE | Standard | 117.9717691 |
| Jul | SE | Standard | 108.6942127 |
| Aug | SE | Standard | 96.97274175 |
| Sep | SE | Standard | 89.70038184 |

*Figure 3 Alternative MPR Unit Sales Calculation*

*Figure 4 Alternative Structure MPR2*

With the MSP structure, the participants had no suggested representation and they had a variety of structures. We identified six different structures, shown in the following figures.

*Figure 5 Structure MSP1*

*Figure 6 Structure MSP2*

*Figure 7 Structure MSP3*

*Figure 8 Structure MSP4*

*Figure 9 Structure MSP5*

*Figure 10 Structure MSP6*

Structures MSP1 and MSP3 are similar, and so are MSP2 and MSP4. We can also note the similarity of structures MPR2 and MSP6: they both perform all the calculations in a single column.

### 4.2 Four-dimensional variables

Seven participants avoided calculating the two four-dimensional variables, `MSPR Unit Sales` and `MSPR Variable Cost`. The 10 other participants had 6 different structures, shown in the following figures. StructuresMSPR1 and MSPR4 were each used by 3 participants, all the others were used by a single participant.

Proceedings of the EuSpRIG 2019 Conference "Spreadsheet Risk Management" ISBN : 978-1-905404-56-8
Copyright © 2019, EuSpRIG European Spreadsheet Risks Interest Group (www.eusprig.org) & the Author(s)
Page 5/12

*Figure 11 Structure MSPR1*

*Figure 12 Structure MSPR2*

*Figure 13 Structure MSPR3*

*Figure 14 Structure MSPR4*

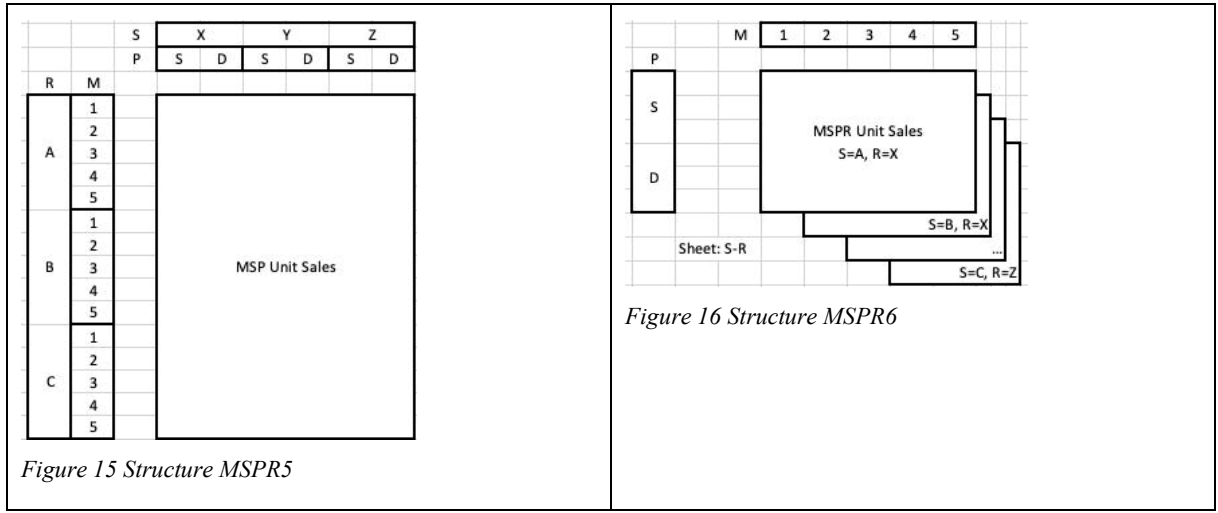

*Figure 15 Structure MSPR5*

*Figure 16 Structure MSPR6*

### 4.3 Database approach

Some participants used a strategy we call the Database approach. This approach consist of using structure MSPR4 and calculating all the problem variables in each row of the table. The columns M, S, R and P collectively form the primary key of the table and, separately, serve as foreign keys to form relations with the other dimension sets. The resulting table was then used with Excel database tools like Power Query, Pivot Tables or SUMIF functions to calculate all the aggregate variables required in the Interface sheet.

Using Power Pivot and Pivot Tables requires a manual recalculation after changing the `Base Price` input variable, and can be inconvenient for users who want to do scenario analysis with Excel's What-If Tables.

## 5 Discussion

As noted above, some participants avoided creating 3 or 4 dimensional variables that were not required in the Interface sheet, even though they were in the supplied Formula List. As a result, their formulas tended to be much more complex than those using the multi-dimensional variables. For example, the formula to calculate the `MPR Unit Sales` using `MSPR Unit Sales` is described in the Formula List as:

`MPR Unit Sales = SUM(MSPR Unit Sales)`

It is implemented as

`=SUM(J51:J54)`

and

`=SUMIFS(tblProcess[MSPR Unit Sales], tblProcess[Product Name],$E$4,tblProcess[Region Name],$G3,tblProcess[Month Name],H$1)`

and

`=SUMIFS(Sales[UnitSales],Sales[Region],$F3,Sales[Month],G$1,Sales[TypeAbbrev],$D$4)`

by participants who used a MSPR structure,

and as

`=SUMPRODUCT((G$31:G$34)*($W$31:$W$34)*($Y$31:$AC$34)*($Y$30:$AC$30=$F3))`

and

`{=SUMPRODUCT(Data!$F$29:$F$32*Data!$B$11:$B$14*TRANSPOSE(Data!$J5:$M5)*TRANSPOSE(OFFSET(Data!$J$14:$M$14,COLUMNS($G$1:G$1),)))}`

and

```
=SUMPRODUCT(OFFSET(Data!$J$5,MATCH($F3,Data!$I$5:$I$9,0)-
1,0,1,4),OFFSET(Data!$T$15,MATCH(G$1,Data!$I$15:$I$26,0)-1,0,1,4))
```

by participants who did not use a MSPR structure.

Some participants also mixed variables of different dimension sets in the same calculation areas. This is often done to calculate aggregate values and presents a visually pleasing table view, as shown in Figure 17 where the body of the table calculates variable `MSP Unit Sales` and the column labelled Total calculates `MS Unit Sales`. This organization makes it hard to copy formulas from left to right, and is a potential error risk during future maintenance.

| Month | Government | | | Military | | |
|---|---|---|---|---|---|---|
| | Standard | Deluxe | Total | Standard | Deluxe | Total |
| Jan | 133 | 72 | 205 | 60 | 179 | 238 |
| Feb | 148 | 80 | 227 | 67 | 201 | 268 |
| Mar | 177 | 96 | 273 | 74 | 223 | 298 |

*Figure 17 Mixing variables of different dimension sets*

Two participants used worksheets to represent one or two dimensions. One used *bookends* sheets to simplify the use of the SUM function (see Figure 18 and structure MSPR6).

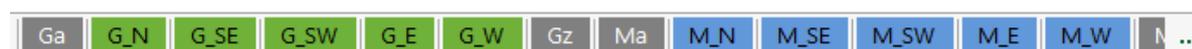

*Figure 18 Modeling dimensions with sheets*

The formula to calculate MPR Unit Sales is simply `=SUM(Ga:Gz!G3)`. The bookend sheets (Ga, Gz, Ma, Mz, etc.) are empty, and one would only need to add a sheet between them to have it used in the SUM. Thus, adding a Sector-Region sheet in the proper place would not require any change in the aggregate formulas.

Participants using the Database approach usually have the simplest formulas for aggregated variables. We expect that they will be the easiest to maintain.

## 6 Conclusion

The analysis of the Multi-dimensional Spreadsheet Challenge submissions has given us insight on the different implementation strategies used by seasoned spreadsheet developers. Given the same problem and its solution, developers had very different results. This lack of standardization may explain the why it is hard to understand and maintain somebody else's spreadsheet.

This study should open the door for more research on complexity and on error reduction. All previous research on complexity has been done on the analysis of existing spreadsheets without examining their original requirements. By ignoring the nature of the problem, they implicitly assume that either the nature of problem itself is not important, or that all problems are equivalent.

An interesting experiment would be to submit the spreadsheets to different developers, along with some requirement changes, and examine the process they follow to understand it and modify it to satisfy the new requirements. We could then infer if a multi-dimensional structure leads to a better understanding and reduces maintenance errors.

Since the solution was supplied in this research, another future research project could just provide a problem statement to participants and examine their resulting spreadsheet with its documentation. This would more closely represent a real-world situation. We expect this to be hard to set up such an experiment outside of an academic setting.

## Appendix A – Multidimensional Spreadsheet Challenge Kit

This section presents the material that was sent to the participants of the Multidimensional Spreadsheet Challenge in November 2018. It consists of the problem statement, the Formula Diagram, the Formula List and two screen captures showing the result for two different inputs.

# Acme Techno Widgets Company

The Acme Techno Widgets Company (ATW) produces and sells widgets. Its salesforce is assigned to four major sectors: Government, military, education and private. It produces two products, the Standard widget and the Deluxe widget.

Market research has established that the annual demand for widgets depends on each sector's Standard widget price. The Pricing Director explains:

*We start by setting a global base price. Then, for each sector, we tell our salesforce that they can offer a rebate. For instance, we offer a 70% rebate to the education sector and it's 10% for the private sector because purchases are usually made by researchers with limited funds. The military sector gets a 20% rebate and the government 40%. This is not made public: all our price lists show the base price, but our clients in each sector are aware of the rebate they can get.*

*Each sector reacts differently to a change of price. We consulted with a market research expert and she came up with multiple demand functions, one for each sector. The demand function estimates a sector's annual demand for a given base price. The demand function has the form $B/Price^A$. The parameters $A$ and $B$ are different for each sector, and $Price$ is the sector's price, after the rebate. This table shows the values the expert gave us:*

| Sector | Government | Military | Private Sector | Education |
|---|---|---|---|---|
| **Rebate Percentage** | 40% | 20% | 10% | 70% |
| **DemParA** | 3.59 | 3.46 | 3.18 | 4.11 |
| **DemParB** | 22000000000 | 22000000000 | 22000000000 | 22000000000 |

*The price of the Deluxe widget is 45% higher than the Standard widget.*

The Sales Manager explains the sales pattern:

*The annual demand of each Sector is split between the Standard and Deluxe products, but the distribution is very different in each sector. For instance, in the education sector, with its limited funds, the split is 80%-20% and it is 25%-75% in the military sector. I guess these guys always go for the best, and they have higher budgets. The distribution is 65%-35% for the government sector and 40%-60% for the private sector. The ratios are then applied to the sector's annual demand to get the annual demand by product.*

*Another interesting pattern is the distribution of sales during the year. We noticed that our clients buy more just before the end of their fiscal year, when some want to spend their budget surpluses, and the beginning, when others have new funds allotted. Each sector has a different pattern, and we noticed that it is pretty stable year after year.*

|  | Government | Military | Private Sector | Education |
|---|---|---|---|---|
| Jan | 9% | 8% | 12% | 6% |
| Feb | 10% | 9% | 11% | 8% |
| Mar | 12% | 10% | 9% | 9% |
| Apr | 12% | 12% | 7% | 10% |
| May | 11% | 13% | 6% | 12% |
| Jun | 9% | 11% | 4% | 12% |
| Jul | 7% | 9% | 5% | 11% |
| Aug | 6% | 7% | 6% | 9% |
| Sep | 5% | 6% | 8% | 7% |
| Oct | 5% | 4% | 9% | 6% |
| Nov   | 6%   | 5%   | 11%  | 5%   |
| Dec   | 8%   | 6%   | 12%  | 5%   |
| Total | 100% | 100% | 100% | 100% |

*Sales to a sector are not uniformly distributed by region. For example, there are more universities in the South-West than in the West. The following table shows the distribution of a sector's sales by region. With it, we can calculate the expected monthly sales per product per region, which helps our Logistics Department do its planning.*

|       | Government | Military | Private Sector | Education |
|-------|------------|----------|----------------|-----------|
| N     | 25%        | 52%      | 22%            | 24%       |
| SE    | 18%        | 13%      | 21%            | 15%       |
| SW    | 18%        | 18%      | 17%            | 32%       |
| E     | 22%        | 0%       | 25%            | 17%       |
| W     | 17%        | 17%      | 15%            | 12%       |
| Total | 100%       | 100%     | 100%           | 100%      |

The costs of producing a widget are $48 and $72 for the Standard and the Deluxe widget respectively. The monthly fixed costs for this year are $20000. Delivery costs depend solely on the region and are shown in this table:

| Region             | North   | South-East | South-West | East   | West    |
|--------------------|---------|------------|------------|--------|---------|
| **Unit Delivery Cost** | $10.25  | $9.73      | $9.58      | $8.26  | $11.02  |

The company CEO wants to see the following results:
- The monthly unit sales per product per region.
- The monthly sales amount and unit sales per product.
- The monthly unit sales and profit.
- The total profit.

### Acme TechnoWidget Company Formula Diagram

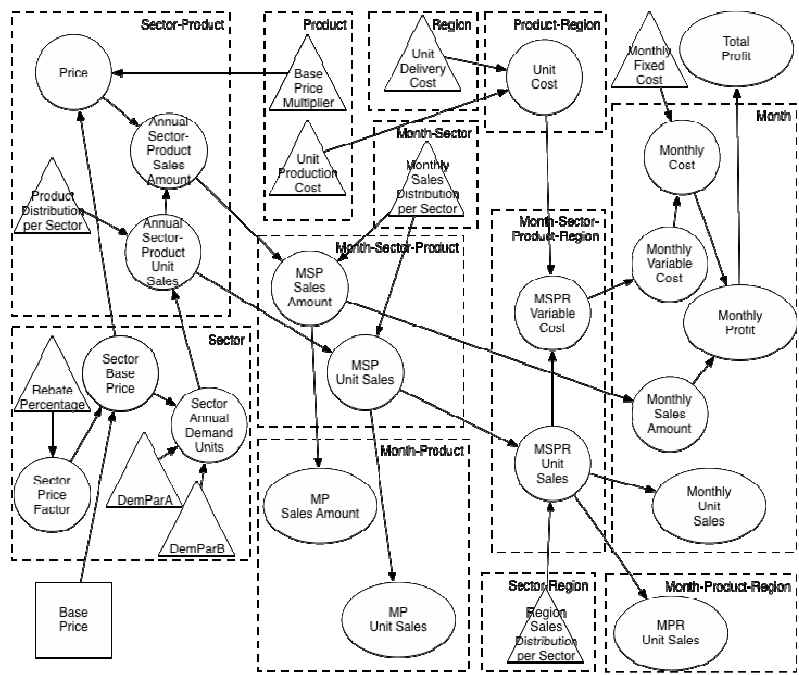

**Acme TechnoWidget Company Formula List**

| Var No | Variable | Type | Dimension Set | Value / Formula |
|---|---|---|---|---|
| 1 | Base Price | Input | | $100 |
| 2 | Base Price Multiplier | Data | Product | (1, 1.45) |
| 3 | Unit Production Cost | Data | Product | list of values |
| 4 | Rebate Percentage | Data | Sector | list of values |
| 5 | Sector Price Factor | Calculated | Sector | 1-Rebate Percentage |
| 6 | Sector Base Price | Calculated | Sector | Base Price * Sector Price Factor |
| 7 | DemParA | Data | Sector | list of values |
| 8 | DemParB | Data | Sector | list of values |
| 9 | Sector Annual Demand Units | Calculated | Sector | DemParA*DemParB^-Sector Base Price |
| 10 | Unit Delivery Cost | Data | Region | list of values |
| 11 | PR Unit Cost | Calculated | Product-Region | Unit Production Cost + Unit Delivery Cost |
| 12 | Product Distribution per Sector | Data | Sector-Product | list of values |
| 13 | Annual Sector-Product Unit Sales | Calculated | Sector-Product | Sector Annual Demand Units * Product Distribution per Sector |
| 14 | Price | Calculated | Sector-Product | Sector Base Price * Base Price Multiplier |
| 15 | Annual Sector-Product Sales Amount | Calculated | Sector-Product | Annual Sector-Product Unit Sales * Price |
| 16 | Region Sales Distribution per Sector | Data | Sector-Region | list of values |
| 17 | Monthly Sales Distribution per Sector | Data | Month-Sector | list of values |
| 18 | MSP Unit Sales | Calculated | Month-Sector-Product | Annual Sector-Product Unit Sales * Monthly Sales Distribution per Sector |
| 19 | MSP Sales Amount | Calculated | Month-Sector-Product | Annual Sector-Product Sales Amount * Monthly Sales Distribution per Sector |
| 20 | MSPR Unit Sales | Calculated | Month-Sector-Product-Region | MSP Unit Sales * Region Sales Distribution per Sector |
| 21 | MSPR Variable Cost | Calculated | Month-Sector-Product-Region | MSPR Unit Sales * PR Unit Cost |
| 22 | Monthly Variable Cost | Calculated | Month | SUM(MSPR Variable Cost) |
| 23 | Monthly Unit Sales | Output | Month | SUM(MSPR Unit Sales) |
| 24 | Monthly Sales Amount | Calculated | Month | SUM(MSP Sales Amount) |
| 25 | Monthly Fixed Cost | Data | | $20000 |
| 26 | Monthly Costs | Calculated | Month | Monthly Fixed Cost + Monthly Variable Cost |
| 27 | Monthly Profit | Calculated | Month | Monthly Sales Amount - Monthly Costs |
| 28 | MPR Unit Sales | Output | Month-Product-Region | SUM(MSPR Unit Sales) |
| 29 | MP Unit Sales | Output | Month-Product | SUM(MSP Unit Sales) |
| 30 | MP Sales Amount | Output | Month-Product | SUM(MSP Sales Amount) |
| 31 | Total Profit | Output | | SUM(Monthly Profit) |

## References

Bregar, A. (2004). Complexity Metrics for Spreadsheet Models. Proc EuSpRIG.

EuSpRIG. (2019, 04 01). *EuSpRIG discussion group*. Retrieved from http://groups.yahoo.com/group/eusprig

FAST Standard Organisation. (2016, June). *The FAST Standard.* Retrieved 04 01, 2019, from http://www.fast-standard.org/wp-content/uploads/2016/06/FAST-Standard-02b-June-2016.pdf

Hermans, F., Pinzger, M., & van Deursen, A. (2014). Detecting and Refactoring Code Smells in Spreadsheet Formulas. *Empirical Software Engineering* (April).

LinkedIn. (2019, 04 01). *Excel Developers*. Retrieved from LinkedIn: https://www.linkedin.com/groups/58704/

Mireault, P. (2018). Structured Spreadsheet Modelling and Implementation with Multiple Dimensions - Part 2: Implementation. Proc. *EuSpRIG.* London.

Raffensperger, J. F. (2003). New Guidelines for Spreadsheets. *International Journal of Business and Economics, 2*(2), 141-154.

Reschenhofer, T., Waltl, B., & Matthes, F. (2016). A Conceptual Model for Measuring the Complexity of Spreadsheets. London: Proc. EuSpRIG.